# Measurements of 1/f noise in Josephson junctions at zero voltage: Implications for decoherence in superconducting quantum bits

Michael Mück[a)] and Matthias Korn
*Institute of Applied Physics, Justus-Liebig-Universität Giessen, 35392 Giessen, Germany*

C. G. A. Mugford and J. B. Kycia
*Department of Physics and the Institute for Quantum Computing, University of Waterloo, Waterloo, Canada, ON N2L 3G1*

John Clarke[b)]
*Department of Physics, University of California, Berkeley, California 94720-7300*

Critical current fluctuations with a $1/f$ spectral density ($f$ is frequency) are potentially a limiting source of intrinsic decoherence in superconducting quantum bits (qubits) based on Josephson tunnel junctions. Prior measurements of this noise were made at nonzero voltages whereas qubits are operated in the zero voltage state. We report measurements of $1/f$ noise in a dc superconducting quantum interference device first, coupled to a resonant tank circuit and operated in a dispersive mode at zero voltage, and, second, operated conventionally with a current bias in the voltage regime. Both measurements yield essentially the same magnitude of critical current $1/f$ noise. © 2005 American Institute of Physics. [DOI: 10.1063/1.1846157]

There is substantial interest in experiments demonstrating coherent superpositions of quantum states[1–8] in superconducting circuits. These include a superconducting loop interrupted by one[1] or three[2] Josephson tunnel junctions, the Cooper pair box,[3,4] "quantronium,"[5] and the current-biased Josephson junction.[6–8] A major challenge is to identify mechanisms leading to relaxation and dephasing. In addition to the environment, there are at least two intrinsic mechanisms: the motion of weakly pinned flux vortices and the motion of weakly trapped charges. Although the first can be eliminated by making the superconducting films sufficiently narrow,[9] it is not obvious how to inhibit moving charges. Charge motion produces decoherence through both charge fluctuations in capacitive elements and fluctuations in the coupling energy of Josephson junctions, which gives rise to fluctuations in the critical current, $I_0$.[10–12] In the currently accepted picture,[13–16] a given charge either tunnels or is thermally activated between two states, thereby producing a random telegraph signal. A superposition of such independent processes with a range of characteristic times produces noise with a $1/f$ spectral density ($f$ is frequency).[17] The dephasing effects of critical current $1/f$ noise have been calculated by Martinis et al.[18] and Van Harlingen et al.[19]

In prior measurements,[10–12] low-$T_c$ Josephson junctions were resistively shunted and operated at nonzero voltage. Fluctuations in critical current were inferred from either fluctuations in the voltage across the junction, in the case of current bias, or in the current through the junction, in the case of voltage bias. On the other hand, the superposition of quantum states in qubits always takes place in the *zero-voltage* regime. Given the difficulty in understanding the dependence of the critical current noise on voltage,[19] it is of considerable interest to determine the noise at zero voltage. In this Letter we report measurements of the $1/f$ noise in a dc superconducting quantum interference device (SQUID) at zero voltage, and compare the results with those measured on the same device in the voltage state.

We first consider a dc SQUID operated at zero voltage in a way analogous to the rf SQUID in the dispersive regime[20] (inset, Fig. 1). A flux $\Phi$ applied to the SQUID gives rise to a circulating current $J$ around the loop, which in turn modifies the inductance of the junctions. We relate fluctuations in the critical currents to an equivalent flux noise. For a dc SQUID with inductance $L$, identical junctions, and zero external bias current, the phase differences $\delta_1$ and $\delta_2$ across the two junctions are related to $J$ by $\delta_1 = \sin^{-1}(-J/I_0)$ and $\delta_2 = \sin^{-1}(J/I_0)$. Equating the phase change $\delta_1 - \delta_2$ around the loop to $2\pi\Phi_T/\Phi_0$, where $\Phi_T = (\Phi - LJ)$ is the total enclosed flux, and noting that $\delta_2 = -\delta_1$ we find

$$J = I_0 \sin[\pi(\Phi - LJ)/\Phi_0]. \qquad (1)$$

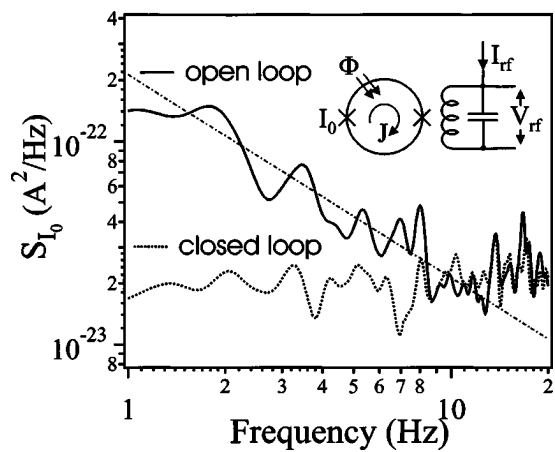

FIG. 1. Spectral densities of critical current noise of the dc SQUID operated in the dispersive mode (inset). The solid line is for open-loop operation with no flux modulation; a line with slope -1 has been drawn through the data. The dotted line is for closed-loop operation with flux modulation.

[a)]Electronic mail: michael.mueck@ap.physik.uni-giessen.de
[b)]Electronic mail: jclarke@physics.berkeley.edu

Although at first sight Eq. (1) might suggest that $J$ is periodic in $2\Phi_0$, there is a "hidden periodicity":[21] even for arbitrarily small values of $\beta_L \equiv 2LI_0/\Phi_0$, the SQUID can make a transition to a neighboring solution of Eq. (1) in a range of applied flux that increases with $\beta_L$. The probability of this transition occurring between two wells in the two-dimensional potential landscape, characterized by $\delta_1$ and $\delta_2$, depends sensitively on $\Phi$. Thus, when $\Phi$ is swept from zero to a sufficiently high value and back to zero, the SQUID can undergo two transitions, in general at different values of $\Phi$. The ensuing hysteresis loop gives rise to dissipation, thus modifying the quality factor $Q$ of a rf tank circuit to which the SQUID is coupled. Fortunately, as we describe later in the discussion of the experimental results, one can distinguish experimentally between dissipative and dispersive behavior, and for the moment we assume that the SQUID does not undergo such transitions.

A fluctuation $\delta I_0$ in the critical current of one of the junctions at constant applied flux produces a fluctuation in the circulating current

$$\delta J(\delta I_0) = \frac{\delta I_0 \sin[\pi(\Phi - LJ)/\Phi_0]}{1 + (\pi L I_0/\Phi_0)\cos[\pi(\Phi - LJ)/\Phi_0]}. \quad (2)$$

We can also obtain an expression for the fluctuation in the circulating current due to a fluctuation in flux $\delta\Phi$ at constant critical current:

$$\delta J(\delta\Phi) = \frac{\delta\Phi(\pi I_0/\Phi_0)\cos[\pi(\Phi - LJ)/\Phi_0]}{1 + (\pi L J_0/\Phi_0)\cos[\pi(\Phi - LJ)/\Phi_0]}. \quad (3)$$

Writing Eqs. (2) and (3) in terms of the spectral densities of the circulating current noise, critical current noise and flux noise, $S_J(f), S_{I_0}(f)$, and $S_\Phi(f)$, respectively, we equate the resulting expressions for $S_J(f)$ to find

$$\frac{S_\Phi(f)}{\Phi_0^2} = \frac{2 S_{I_0}(f)}{\pi^2 I_0^2} \tan^2[\pi(\Phi - LJ)/\Phi_0]. \quad (4)$$

We have included a factor of two to account for the uncorrelated fluctuations in the two critical currents. For given values of $LI_0/\Phi_0$ and $\Phi$ one can solve Eq. (1) numerically for $J$ to find an explicit relation between $S_\Phi(f)/\Phi_0^2$ and $S_{I_0}(f)/I_0^2$.

In the case of a current-biased dc SQUID operated conventionally at a nonzero voltage, incoherent fluctuations in the two critical currents contribute to fluctuations in the current–voltage characteristics in two ways.[22] In-phase components produce fluctuations in the voltage across the SQUID that can be eliminated with flux modulation. Out-of-phase fluctuations produce a fluctuating current around the SQUID loop, and thus a fluctuating flux; these can be eliminated by a combination of flux modulation and bias current reversal. When the SQUID is operated with flux modulation and without bias current reversal, for typical parameters the out-of-phase contribution to the flux noise is[22] $S_\Phi(f)/\Phi_0^2 \approx 0.1 S_{I_0}(f)/I_0^2$ for a flux bias of $(2n+1)\Phi_0/4$ ($n$ is an integer).

Our measurements were made on a dc SQUID with Nb-Al-Al$_x$O$_y$-Nb tunnel junctions grown on an oxidized Si wafer and patterned with conventional photolithography. The 4-$\mu$m$^2$ junctions were defined by windows in SiO. After briefly ion milling the 100-nm-thick Nb base layer, we deposited a 10-nm-thick Al film, oxidized it *in situ* and deposited the 80-nm Nb counter electrode.[11] After patterning and ion milling this layer, we deposited and patterned a 25-nm-thick Pd film to form resistive shunts, each with resistance $R \approx 30\ \Omega$. The maximum critical current ($2I_0$) was 0.9 $\mu$A and the estimated loop inductance 300 pH; thus, $\beta_L \approx 0.13$. All measurements were performed with the SQUID immersed in liquid helium and surrounded with a superconducting shield.

To perform the zero-voltage measurements, we coupled the SQUID inductively to a coil of Nb wire, with an inductance of about 0.13 $\mu$H, in parallel with a 1.5 pF (ceramic) capacitor. The tank circuit was driven off resonance with a 360-MHz current $I_{\rm rf}$ of fixed amplitude, and the voltage $V_{\rm rf}$ across the tank circuit was measured using a cold (4.2 K) low-noise amplifier. After amplification, the 360-MHz signal was demodulated; the demodulated signal was calibrated with a small flux change applied to the SQUID at a flux bias near $(2n+1)\Phi_0/4$. Fluctuations in the junction critical currents modulated the SQUID inductance and thus the resonant frequency of the tank circuit, generating fluctuations in the demodulated signal. We note that when the tank circuit was driven on resonance, there was no discernible change (not more than 1% of the maximum change off resonance) in the rf voltage when we varied the flux through the SQUID. This finding confirms that the SQUID was operating overwhelmingly in the dispersive mode, and that any transitions that could produce dissipation were rare.

We measured the flux noise of the SQUID and converted it to critical current noise as follows. For $\beta_L = 0.13$ and $\Phi = (2n+1)\Phi_0/4$, Eq. (1) yields $J = 0.6 I_0$, and from Eq. (4) we find $S_\Phi(f)/\Phi_0^2 \approx 0.12 S_{I_0}(f)/I_0^2$. We note that, given the inaccuracy of noise measurements, this is essentially the same result as that calculated for the dc SQUID operated in the voltage regime. Figure 1 shows the inferred spectral density of the critical current noise in the SQUID measured in two different ways. First, the flux noise was measured open loop as described above; the noise exhibits a $1/f$ power spectrum. However, *a priori* this noise could arise from either critical current fluctuations or from the motion of flux vortices. Second, in addition to the 360-MHz modulation, the flux in the SQUID was modulated at 100 kHz with a peak-to-peak amplitude of $\Phi_0/2$. The 100-kHz voltage was demodulated with a lock-in detector, integrated, and coupled via a resistor to the inductor of the tank circuit to flux lock the SQUID. This method greatly reduces $1/f$ noise due to critical current fluctuations,[23] but does not affect the $1/f$ noise due to flux motion. We see in Fig. 1 that the $1/f$ noise is eliminated by this double modulation technique, and thus unequivocally arises from critical current fluctuations rather than flux motion. The residual white noise was due predominantly to intrinsic noise in the SQUID. We note that the use of flux modulation to eliminate $1/f$ critical current noise in Nb rf SQUIDs at frequencies down to less than 1 Hz is well established.[23]

Subsequently, we measured the $1/f$ noise in the same device operated in the conventional manner. The SQUID was biased with a constant current $I_B$, and flux modulated at 100 kHz with a peak-to-peak amplitude of $\Phi_0/2$ (inset, Fig. 2). The alternating voltage $V$ across the SQUID was amplified by a room-temperature transformer and a low-noise amplifier, demodulated with a lock-in detector, integrated, and coupled via a resistor to an inductor coupled to the SQUID to form a flux-locked loop. The output voltage was calibrated with a small flux change applied to the SQUID. This mode of operation rejects $1/f$ fluctuations in the two junctions that are in phase,[22] but does not eliminate fluctuations that are out of phase and that induce a fluctuating current around the SQUID loop; furthermore, $1/f$ noise due to flux motion is not reduced. The resulting $1/f$ spectral density of the critical current fluctuations is shown in Fig. 2. The noise was remeasured with bias current reversal: in addition to the 100 kHz flux modulation, the bias current was reversed at 3 kHz, and simultaneously the phase of the lock-in detector was shifted by $\pi$. This technique eliminates out-of-phase fluctuations in the critical current but does not affect fluctuations due to flux motion.[22] Figure 2 shows that bias current reversal eliminates the $1/f$ noise measured in the absence of reversal; thus, the observed $1/f$ noise arises from critical current fluctuations. As in the dispersive mode, the white noise was due largely to the intrinsic noise of the SQUID. Comparing the spectral densities of the $1/f$ critical current noise in Figs. 1 and 2, we see that, to within the scatter of the data, they have the same magnitude.

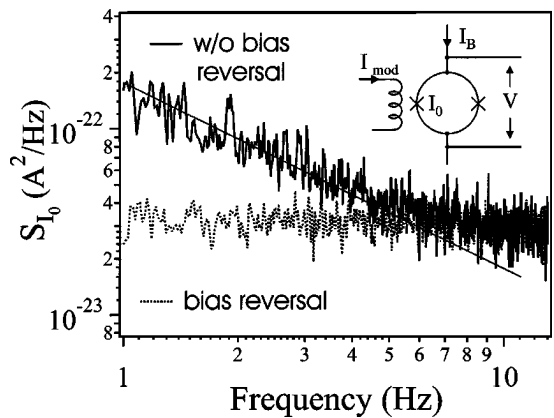

FIG. 2. Spectral densities of critical current noise of the dc SQUID operated with current bias at nonzero voltage (inset) in a flux-locked loop. The solid line is for operation with flux modulation only; a line with slope -1 has been drawn through the data. The dashed line is for operation with flux modulation and bias current reversal.

In summary, we have demonstrated that the $1/f$ noise in a dc SQUID due to fluctuations in the critical currents has the same magnitude measured at zero voltage as in the voltage regime. Thus, the levels of critical current $1/f$ noise measured by numerous authors at nonzero voltages should pertain to qubits operated at zero voltage. The dispersive method of determining $1/f$ noise allows one to extend measurements to low temperatures without raising concerns about dissipation in the shunt produced by the bias current, which can raise the effective junction temperature substantially above the bath temperature.[24]

The authors are indebted to Britton Plourde and Timothy Robertson for helpful discussions. This work was supported by the Air Force Office of Scientific Research under Grant No. F49-620-02-1-0295, by the Army Research Office under Grant No. DAAD-19-02-1-0187, with funds provided by the Advanced R&D Activity (ARDA) of the National Security Agency, by the National Science Foundation under Grant No. EIA-020-5641 (Berkeley), and by the Natural Sciences and Engineering Council and the Canada Foundation for Innovation (Waterloo).